\documentclass[showpacs,superscriptaddress,amsmath,amssymb,aps,preprint]{revtex4}
\usepackage{graphicx}
\usepackage{bm}
\usepackage[utf8]{inputenc}
\usepackage{lmodern}
\begin{document}
\title{Exact solution to neutrino-plasma two-flavor dynamics} 
\author{F. Haas}
\affiliation{Departamento de F{\'i}sica, Universidade Federal do Paran\'a, 81531-990, Curitiba, Paran\'a, Brazil}
\author{J. T. Mendon\c{c}a}
\affiliation{IPFN, Instituto Superior T\'ecnico, 1049-001 Lisboa, Portugal}
\begin{abstract}
It is shown that the two-flavor neutrino oscillation equations admit an exact analytic solution for arbitrarily chosen normalized electron neutrino population, provided the electron plasma density is adjusted in a certain way. The associated formula for the electron plasma density is applied to the cases of exponentially decaying or oscillating electron neutrino populations. 
\end{abstract}
\maketitle
\section{Introduction}
The energy exchange between neutrino beams and plasma collective modes can be a crucial mechanism e.g. for shocks in type II supernovae \cite{Bingham}. The associated neutrino charge coupling \cite{Serbeto} leads to kinetic effects such as neutrino Landau damping \cite{Silva}, as well as to the generation of quasi-static magnetic fields \cite{Shukla1} The orthodox approach to the neutrino-plasma interaction problem is to assume specific medium properties, and then to solve the dynamical equations, either in approximate or numerical forms. In this respect, one can have sinusoidal variations of the electron density \cite{Kneller}, 
\cite{Koike}, \cite{Krastev}, \cite{Schafer}, general time-dependent media \cite{Hollenberg}, stochastic backgrounds \cite{Torrente}, \cite{Benatti} as well as instabilities due to electron density ripples 
\cite{Shukla2}. In an inverse way, in the present work a certain electron density profile is assumed, and then the corresponding medium properties are unveiled. The procedure is restricted to two-flavor neutrino populations. No further approximations are needed. 

The work is organized as follows. Section II describes the general method, leading to Eq. (\ref{e8}), the central result of the paper. Section III briefly discuss the cases of exponentially decaying or oscillating electron neutrino populations. Section IV is reserved to final remarks. 
 
\section{Exact solution}\label{sec2}
The equations for neutrino-flavor oscillations in a plasma are well known \cite{Raffelt} and we present them in the form 
\begin{equation}
\label{e1}
\dot{P}_1 = - \Omega(t) P_2 \,, \quad \dot{P}_2 = \Omega(t) P_1 - \Omega_{0} P_3 \,, \quad \dot{P}_3 = \Omega_{0} P_2 \,,
\end{equation}
where ${\bf P} = (P_1,P_2,P_3)$ is the three-dimensional flavor polarization vector, such that the density
matrix can be written as
\begin{equation}
\label{e2}
\rho = \frac{N_0}{2}(1 + {\bf P}\cdot{\boldsymbol{\sigma}}) \,,
\end{equation}
using the total neutrino number $N_0 = N_e + N_\mu$ and the Pauli matrices $\boldsymbol{\sigma} = (\sigma_{x},\sigma_{y},\sigma_{z})$, with $N_{e,\mu}$ being the electron (muon) neutrino populations. In Eq. (\ref{e1}), 
\begin{equation}
\label{e3}
\Omega(t) = \omega_{0}(\cos 2\theta_0 - \xi(t)) \,, \quad \Omega_0 = \omega_{0}\sin 2\theta_0 \,,
\end{equation}
where we have introduced the characteristic oscillation frequency $\omega_0 = \Delta m^2/2E$, with $\Delta m^2 = m_{2}^2 - m_{1}^2$ being  the square mass difference between mass eigenstates and $E$ the energy associated to the neutrino Dirac spinor, while $\theta_0$ is the pertinent mixing angle. Finally, we have $\xi(t) = \sqrt{2} G_F n_{e}/\omega_0$ being the coupling function between the neutrino and the embedding plasma medium, where $G_F$ is the Fermi constant and $n_e$ the electron plasma density. In our analysis, it is important to keep in mind that $P_3 = (N_e - N_{\mu})/N_0$. 

From the first and the last equations in Eq. (\ref{e1}) we get 
\begin{equation}
\label{e4}
\Omega = - \frac{\dot{P}_1}{P_2} \,, \quad P_2 = \frac{\dot{P}_3}{\Omega_0} \,.
\end{equation}
Substituting the results shown in Eq. (\ref{e4}) into the mid equality in Eq. (\ref{e1}) and integrating once yields
\begin{equation}
\label{e5} 
I = \dot{P}_{3}^2 + \Omega_{0}^2 (P_{3}^2 + P_{1}^2) = \Omega_{0}^2 \,,
\end{equation}
where $I$ is a constant of motion, $dI/dt = 0$. The last equality in Eq. (\ref{e5}) follows from $\dot{P}_3 = \Omega_{0} P_2$ and the normalization condition, $|{\bf P}| = 1$. Our central result comes from the fact that Eq. (\ref{e5}) can be solved up to a sign choice for $P_1$ in terms of $P_3$, or 
\begin{equation}
\label{e6}
P_1 = \pm \frac{(\Omega_{0}^2 - \dot{P}_{3}^2 - \Omega_{0}^2 P_{3}^2)^{1/2}}{\Omega_0} \,.
\end{equation}
Correspondingly, using Eqs. (\ref{e4}) and (\ref{e6}) we find
\begin{equation}
\label{e7}
\Omega = \pm \frac{\ddot{P}_3 + \Omega_{0}^2 P_3}{(\Omega_{0}^2 - \dot{P}_{3}^2 - \Omega_{0}^2 P_{3}^2)^{1/2}} \,.
\end{equation}

Therefore, we have a very simple recipe to generate exact solutions for the two-flavor neutrino-plasma oscillation equations. Instead of prescribing a given plasma density $n_{e}$ as usual, one can start choosing 
$P_{3}$, which is interpreted as the normalized difference between neutrino flavor populations. Afterward, Eqs. (\ref{e6}) and the last in Eq. (\ref{e4}) gives resp. the coherences $P_{1}$ and $P_{2}$. Finally, 
Eq. (\ref{e7}) gives the corresponding $\Omega$, which is linked to the plasma medium properties. To have meaningful solutions at least some requirements should be taken into account, namely $|P_{3}| \leq 1$,  
otherwise one would eventually get negative flavor populations. In addition, $P_3$ should be a double-differentiable function of time.

Alternatively, we can use $P_3 = 2 N_{e}/N_0 - 1$ to express the results in terms of the electron neutrino population. From Eqs. (\ref{e4}), (\ref{e6}) and (\ref{e7}) we get
\begin{equation}
\label{e8} 
P_1 = \pm 2\left(\bar{N}_e - \bar{N}_{e}^2 - \frac{\dot{\bar{N}}_{e}^2}{\Omega_{0}^2}\right)^{1/2} , \quad P_2 = \frac{2\dot{\bar{N}}_{e}}{\Omega_0} \,, \quad \Omega = \pm \frac{\ddot{\bar{N}}_e + \Omega_{0}^2 (\bar{N}_e - 1/2)}{\Omega_0 \left(\bar{N}_e - \bar{N}_{e}^2 - \dot{\bar{N}}_{e}^{2}/\Omega_{0}^2\right)^{1/2}} \,,
\end{equation}
where $\bar{N}_e \equiv N_{e}/N_0$. The results in Eq. (\ref{e8}) compactly represents the basic findings of this work. 

\section{Applications} 
\subsection{Exponentially decaying electron neutrino population}
\begin{figure}
  \centerline{\includegraphics{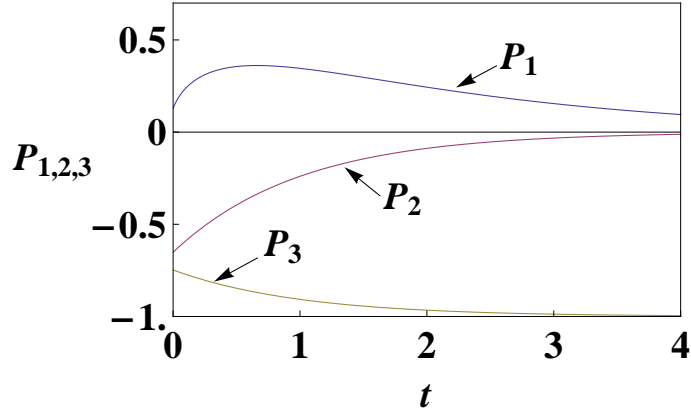}}% Images in 100% size
  \caption{Polarization vector components for an exponentially decaying electron neutrino population, according to Eqs. (\ref{e8}) and (\ref{e9}). Parameters, $\omega_0 = 1, r_0 = 1$, $\sin^{2}2\theta_0 = 0.15$, 
	$\bar{N}_{e}(t_0) \exp(t_{0}/r_{0}) = 0.13$.}
\label{fig1}
\end{figure}

\begin{figure}
  \centerline{\includegraphics{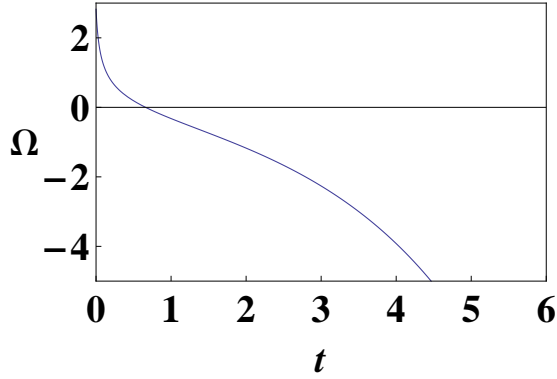}}% Images in 100% size
  \caption{Function $\Omega(t)$ for an exponentially decaying electron neutrino population, according to Eqs. (\ref{e8}) and (\ref{e9}) and the same parameters as in Fig. \ref{fig1}.}
\label{fig2}
\end{figure}

As a first example, consider the case of an exponentially decaying electron neutrino population,
\begin{equation}
\label{e9}
\bar{N}_e = \bar{N}_{e}(t_0) \exp\left(- \frac{t-t_0}{r_0}\right) \,,
\end{equation}
which models the change of the electron number density along the
path of the solar neutrinos moving radially from the central region to the surface of the Sun \cite{Petkov2}, \cite{Petkov1}. In this context $r_0$ is the scale height and $t-t_0$ is the distance traveled by the neutrinos. To have meaningful solutions from Eq. (\ref{e8}) (or, real $P_{1,2,3}$) one should have $\bar{N}_{e}(t_0) \exp(t_{0}/r_{0}) > (1 + 1/\Omega_{0}^2 r_{0}^2)^{-1}$, as can be readily verified. We use stretched time and space variables so that $\omega_0 = 1, r_0 = 1$. Moreover, the mixing angle satisfy $\sin^{2}2\theta_0 = 0.15$, so that $\Omega_0 = 0.39$. Finally, we chose $\bar{N}_{e}(t_0) \exp(t_{0}/r_{0}) = 0.13$, which assures the produced solutions to be non-complex. The resulting polarization vector components are shown in Fig. \ref{fig1}, while $\Omega(t)$ is shown in Fig. \ref{fig2}, with the plus sign chosen in Eq. (\ref{e8}). 
It can be shown that in this case one has the asymptotic dependence $\Omega \propto - \Omega_0 \exp(t/2 r_0)$ when $t \rightarrow \infty$. Evidently, an infinite class of profiles can be generated via the same procedure. One can e.g. consider the case of an oscillating electron neutrino population, discussed in the following.

\subsection{Periodic electron density}
\begin{figure}
  \centerline{\includegraphics{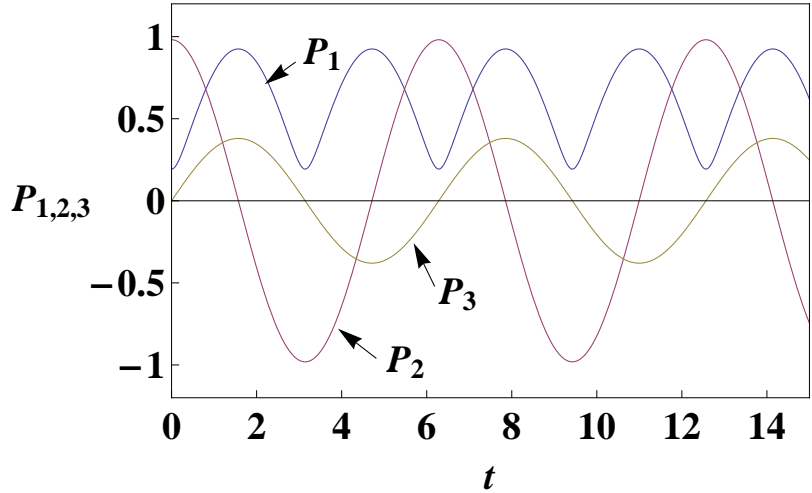}}% Images in 100% size
  \caption{Polarization vector components for an oscillating electron neutrino population, according to Eqs. (\ref{e8}) and (\ref{e10}). Parameters, $\Omega_0 = 0.39, \varepsilon = 0.39, \tilde{\Omega} = 1.0$.}
\label{fig3}
\end{figure}

\begin{figure}
  \centerline{\includegraphics{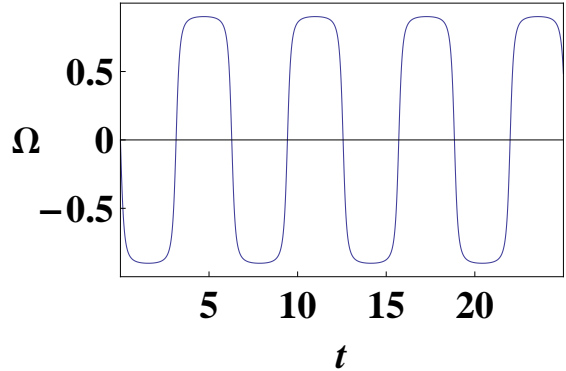}}% Images in 100% size
  \caption{Function $\Omega(t)$ for an oscillating electron neutrino population, according to Eqs. (\ref{e8}) and (\ref{e10}) and the same parameters as in Fig. \ref{fig3}.}
\label{fig4}
\end{figure}

Now consider an initially unpolarized electron neutrino beam, 
\begin{equation}
\label{e10}
\bar{N}_e = \frac{1}{2} + \frac{\varepsilon}{2}\sin\tilde{\Omega}t \,,
\end{equation}
including an amplitude parameter $\varepsilon \geq 0$ and an arbitrary frequency $\tilde{\Omega}$. 
A simple analysis shows that $\varepsilon < {\rm Inf}(1, \Omega_{0}/\tilde{\Omega})$ is the condition to avoid singularities. The corresponding polarization vector components and $\Omega(t)$ function are shown resp. in Figs. \ref{fig3} and \ref{fig4}, for $\Omega_0 = 0.39$ as before and for $\varepsilon = 0.39, \tilde{\Omega} = 1.0$.  

\section{Conclusion} 
In this work the usual route for solving the two-flavor neutrino-plasma oscillation equations has been subverted. Namely, instead of setting a certain electron plasma density and then looking for the polarization vector components, here the third component $P_{3}(t)$ and equivalently the electron neutrino population $N_{e}(t)$ are chosen {\it ab initio}. Consequently, simple formulas for the coherences $P_{1,2}(t)$ are readily found. The necessary condition for the recipe to work is to adjust the function $\Omega(t)$ and hence the electron plasma density $n_{e}(t)$ so that Eq. (\ref{e7}) holds. The results can be expressed in terms of the electron neutrino population only, see Eq. (\ref{e8}). In a sense, our exact neutrino flavor solution has similarities with the celebrated Bernstein-Greene-Kruskal equilibria for the Vlasov-Poisson system 
\cite{Bernstein}, where arbitrarily chosen electrostatic potentials can be constructed provided specific trapped electron distributions are set. 

{\bf Acknowledgments}\\
Fernando Haas acknowledge CNPq (Conselho Nacional de Desenvolvimento Cient\'{\i}fico e Tecnol\'ogico) for financial support. This work is dedicated 
to the memory of Prof. Padma Kant Shukla.

\end{document}